\documentclass[pra,twocolumn,amsmath,amssymb,nofootinbib]{revtex4}

\usepackage{graphicx}
\usepackage{amsthm}

\begin{document}
\title{
Geometrical self-testing of partially entangled two-qubit states}
\author{Satoshi Ishizaka}
\affiliation{Graduate School of Integrated Arts and Sciences,
Hiroshima University,
1-7-1 Kagamiyama, Higashi-Hiroshima 739-8521, Japan}
\date{\today}
%
\begin{abstract}
Quantum nonlocality has recently been intensively studied in connection to
device-independent quantum information processing, 
where the extremal points of
the set of quantum correlations play a crucial role through
self-testing.
In most protocols, the proofs for self-testing rely on the
maximal violation of the Bell inequalities, but there is another known
proof based on the geometry of state vectors to self-test a maximally
entangled state.
We present a geometrical proof in the case of partially entangled states.
We show that, when a set of correlators in the simplest Bell scenario satisfies
a condition, the geometry of the state vectors is uniquely
determined.
The realization becomes self-testable when another unitary
observable exists on the geometry.
Applying this proven fact, we propose self-testing protocols by intentionally
adding one more measurement.
This geometrical scheme for self-testing is superior in that, by using this
as a building block and repeatedly adding measurements, a realization
with an arbitrary number of measurements can be self-tested.
Besides the application, we also attempt to describe nonlocal correlations by 
guessing probabilities of distant measurement outcomes.
In this description, the quantum set is also convex,
and a large class of extremal points is identified by
the uniqueness of the geometry.
\end{abstract}
%
\maketitle
%
\section{Introduction}
\label{sec: Introduction}

It was shown by Bell that the nonlocal correlations predicted by
quantum mechanics are inconsistent with local realism \cite{Bell64a}.
Bell nonlocality, or quantum nonlocality, has attracted many
research interests over the years (see \cite{Brunner14a} for a
review).
Recently, it has been
intensively studied in connection to device-independent quantum information
processing (see \cite{Scarani12a,Supic19a} for reviews),
where the extremal points of the convex set of quantum correlations plays a
crucial role through self-testing.

The correlation that attains the maximal quantum violation of
$2\sqrt{2}$ \cite{Tsirelson80a} in the Clauser-Horne-Shimony-Holt
(CHSH) inequality \cite{Clauser69a} is an extremal point of the quantum
set, for which the quantum realization (state and measurements) 
is unique up to unavoidable local isometry.
This implies that attaining the value of $2\sqrt{2}$ can self-test
the state and the measurements in the Bell experiment \cite{Mayers04a}.
When a realization is a unique maximizer of a Bell inequality,
the realized correlation is a self-testable extremal point.
Although there exist non-exposed extremal points that cannot be a unique
maximizer of any Bell inequality, a correlation is extremal when the
realization is self-testable \cite{Goh18a}.
In this way, self-testability and extremality are intimately connected.
In most protocols, the proofs for self-testing
rely on the maximal violation of the Bell inequalities.
However, even in the simplest Bell scenario (two parties and two binary
measurements on each party), the maximal violation by a partially
entangled state is known for only a few Bell
inequalities \cite{Liang11a,Acin12a,Ramanathan18a,Baccari18a,Wagner18a},
and not many protocols are proposed for self-testing partially entangled
states \cite{Yang13a,Bamps15a,Rabelo12a,Coladangelo17a,Woodhead19a}.

On the other hand, the proof for self-testing in \cite{Wang16a} is fascinating,
because no Bell inequality is used directly.
In the simplest Bell scenario, when marginal probabilities of outcomes are
unbiased, the boundaries of the quantum set are identified
by the Tsirelson-Landau-Masanes (TLM)
criterion \cite{Tsirelson87a,Landau88a,Masanes03a}.
The proof in \cite{Wang16a} relies on the fact that the geometry of the
state vectors is uniquely determined when the TLM
criterion is satisfied (and the anti-commutation relation between
observables is proven on the geometry).
However, this geometrical proof is restricted to the case of
a maximally entangled state by the restriction of the TLM criterion.
In a general case, where an extremal correlation may be realized by
a partially entangled state, the criterion for the identification
has been only conjectured, based on the probabilities of guessing outcomes
of a distant party
(referred as ``guessing probability'' hereafter) \cite{Ishizaka18a}.

In this paper, we present a geometrical proof in the case of partially
entangled states.
We show that, when a set of correlators in the simplest Bell scenario
satisfies a condition, the geometry of state vectors is
uniquely determined.
The realization becomes self-testable when
another unitary observable exists on the geometry to prove anti-commutation
relation.
Applying this proven fact, we propose self-testing protocols by intentionally
adding one more measurement to prove the anti-commutation relation.
This geometrical scheme for self-testing is superior in that, by using this
as a building block and repeatedly adding measurements, a realization
with an arbitrary number of measurements can be self-tested.

Beside applications, efforts have been made to describe
the quantum set having a complicated
structure \cite{Wolfe12a,Donohue15a,Goh16a,Goh18a,Rai19a} in a more
tractable way; some descriptions exist such as
covariance \cite{Pozsgay17a} and entropy \cite{Cerf97a}.
For this purpose, we attempt to describe nonlocal correlations by 
guessing probabilities.
We show that the quantum realizable set is also convex in this description,
and a large class of extremal points is identified by the
uniqueness of the geometry of state vectors.
Moreover, with the help of this extremality, we show that the sufficiency of
the extremal criterion conjectured in \cite{Ishizaka18a} can be
reduced to certifiability of guessing probabilities.

This paper is organized as follows:
In Sec.\ \ref{sec: Preliminaries}, we briefly summarize the preliminaries.
For details, see \cite{Brunner14a,Scarani12a,Supic19a} and the references
therein.
For clarity, we first introduce the description of correlations by
guessing probabilities in Sec.\ \ref{sec: Quantum set in D-space},
and discuss the properties of the quantum set, such as the extremality and
self-testability.
In Sec.\ \ref{sec: Quantum set in C-space}, we investigate the geometrical
properties of realizations in the standard description of correlations.
Finally, as an application, we propose self-testing protocols for
partially entangled states in
Sec.\ \ref{sec: Scheme for self-testing partially entangled state},
whose self-testability is geometrically proven, regardless of the validity
of the conjectured extremal criterion.
A summary is given in Sec.\ \ref{sec: Summary}.

%
\section{Preliminaries}
\label{sec: Preliminaries}

In the simplest Bell scenario, 
Alice (Bob) performs one of two binary measurements on a shared 
state depending on a given random bit $x$ ($y$), and obtains an outcome
$a\!=\!\pm 1$ ($b\!=\!\pm1$).
The properties of a nonlocal correlation
are described by a set of conditional probabilities
$\{p(ab|xy)\}$ referred as a ``behavior'', 
which specifies a point in the probability space.
As $p(ab|xy)\!=\!\frac{1}{4}[1+aC^{A}_x+bC^{B}_y+ab C_{xy}]$
for no-signaling correlations, with $C^{A}_x$ ($C^{B}_y$) being a bias
of the marginal $p(a|x)$ [$p(b|y)$], any no-signaling
correlation can be described by a behavior
$\{C^{A}_x,C^{B}_y,C_{xy}\}$.
Such a behavior specifies a point in the 8-dimensional no-signaling space,
which we denote by the $C$-space.

A behavior $\{C^{A}_x,C^{B}_y,C_{xy}\}$
is realized by quantum mechanics, if and only if
there exist a shared quantum state $|\psi\rangle$ and the observables
$A_x$ ($B_y$) of Alice (Bob), such that $A^{2}_x\!=\!B^{2}_y\!=\!I$,
$C^{A}_x\!=\!\langle \psi |A_x|\psi\rangle$,
$C^{B}_y\!=\!\langle \psi |B_y|\psi\rangle$, and
$C_{xy}\!=\!\langle \psi |A_x B_y|\psi\rangle$.
We use $\langle \cdots \rangle$ as the abbreviation of
$\langle\psi| \cdots |\psi\rangle$.
Any state vector has a real-vector representation
\cite {Tsirelson87a,Acin06a,Vertesi08a}.
For example, when $|\psi\rangle$ is represented by
components as $|\psi\rangle\!=\!(c_0,c_1,\cdots)$ with $c_i\in\mathbb{C}$,
$\vec\psi\!=\!({\rm Re~}c_0,{\rm Im~}c_0,{\rm Re~}c_1,{\rm Im~}c_1,\cdots)$
is a real-vector representation.

The realizable behaviors constitute a convex set in the $C$-space, denoted
by ${\cal Q}_C$.
In the unbiased case where $C^{A}_x\!=\!C^{B}_y\!=\!0$, a behavior belongs
to ${\cal Q}_C$, if and only if the TLM
inequality \cite{Tsirelson87a,Landau88a,Masanes03a}
\begin{eqnarray}
|\tilde C_{00}\tilde C_{01}-\tilde C_{10}\tilde C_{11}|
&\!\!\!\le\!\!\!&\sqrt{(1-\tilde C^{2}_{00})(1-\tilde C^{2}_{01})} \cr
&\!\!\!\!\!\!&+\sqrt{(1-\tilde C^{2}_{10})(1-\tilde C^{2}_{11})},
\label{eq: TLM inequality}
\end{eqnarray}
is satisfied for $\tilde C_{xy}\!=\!C_{xy}$ [together with $p(ab|xy)\!\ge\!0$].

Using the correlators of a behavior $\{C^{A}_x,C^{B}_y,C_{xy}\}$,
let us introduce the quantities $S^{\pm}_{xy}$ given by 
\begin{eqnarray}
S^{\pm}_{xy}&\!\!\!\equiv\!\!\!&\frac{1}{2}\left[J_{xy}
\pm\sqrt{J^{2}_{xy}-4K^{2}_{xy}}\right], \nonumber\\
&&J_{xy}\equiv C_{xy}^{2}-(C^{A}_x)^2-(C^{B}_y)^2+1, \nonumber\\
&&K_{xy}\equiv C_{xy}-C^{A}_x C^{B}_y.
\label{eq: Quantities}
\end{eqnarray}
Suppose that the following holds for a set $\{p_{xy}\}$
\begin{eqnarray}
&&S^{p_{00}}_{00}=S^{p_{01}}_{01}=S^{p_{10}}_{10}=S^{p_{11}}_{11}, \nonumber\\
&&H\equiv\prod_{xy}[(1-S^{p_{xy}}_{xy})C_{xy}-C^{A}_x C^{B}_y]\ge0,
\label{eq: Two-qubit condition}
\end{eqnarray}
where $p_{xy}$ is either '$+$' or '$-$'.
Letting the value of $S_{xy}^{p_{xy}}$ be equal to $\sin^22\chi$, the
following is also introduced: 
\begin{equation}
d^{B}_{x}\equiv(C^{A}_x)^2+\sin^22\chi, \hbox{~~}
d^{A}_{y}\equiv(C^{B}_y)^2+\sin^22\chi.
\label{eq: d quantity}
\end{equation}
Then, to identify the nonlocal extremal points of ${\cal Q}_C$,
the following criterion has been conjectured in \cite{Ishizaka18a}.

{\bf Conjecture 1.}
{\it A nonlocal behavior 
$\{C^{A}_x,C^{B}_y,C_{xy}\}$ is an extremal point of ${\cal Q}_C$, if and only if
Eq.\ (\ref{eq: Two-qubit condition}) holds as
$S^{+}_{00}\!=\!S^{+}_{01}\!=\!S^{+}_{10}\!=\!S^{+}_{11}$,
and Eq.\ (\ref{eq: TLM inequality}) is saturated
for both scaled correlators
$\tilde C_{xy}\!=\!C_{xy}/\sqrt{d^{B}_{x}}$
and  $\tilde C_{xy}\!=\!C_{xy}/\sqrt{d^{A}_{y}}$.}

Note that the fulfillment of Eq.\ (\ref{eq: Two-qubit condition}) for some
$\{p_{xy}\}$ (not necessarily as
$S^{+}_{00}\!=\!S^{+}_{01}\!=\!S^{+}_{10}\!=\!S^{+}_{11}$) is necessary
(and even sufficient in the case of $\sin^22\chi\!<\!1$) for the existence of a
two-qubit realization in the form of
\begin{eqnarray}
A_x&\!\!\!=\!\!\!&\sin \theta^{A}_x \sigma_1 + \cos \theta^{A}_x \sigma_3
, \hbox{~~} B_y=\sin \theta^{B}_y \sigma_1 + \cos \theta^{B}_y \sigma_3,
\nonumber\\
|\psi\rangle&\!\!\!=\!\!\!&\cos\chi|00\rangle+\sin\chi|11\rangle,
\label{eq: Two-qubit realization}
\end{eqnarray}
where $(\sigma_1,\sigma_2,\sigma_3)$ are the Pauli matrices
(but there is no $\sigma_2$ term),
and hence also necessary for the extremality of ${\cal Q}_C$
(see the supplemental material of \cite{Ishizaka18a}).
Note further that the definition of $\theta^{A}_x$ and $\theta^{B}_y$ are
changed from \cite{Ishizaka17a,Ishizaka17aE,Ishizaka18a} for convenience
($\theta^{A}_x\rightarrow\pi/2\!-\!\theta^{A}_x$ and $\theta^{B}_y\rightarrow\pi/2\!-\!\theta^{B}_y$).

Moreover, for a given $\{C^{A}_x,C^{B}_y,C_{xy}\}$,
the quantity $D_{x}^B$ and $D_{y}^A$ (explained later) has a
device-independent upper bound,
which can be obtained by the Navascu\'es-Pironio-Ac\'{\i}n
(NPA) hierarchy \cite{Navascues07a,Navascues08a}, and the following is also
implicitly conjectured in \cite{Ishizaka18a}.

{\bf Conjecture 2.}
{\it When a nonlocal behavior 
$\{C^{A}_x,C^{B}_y,$ $C_{xy}\}$ satisfies the same condition as
Conjecture 1, $d^{B}_x$ and $d^{A}_y$ coincides with the device-independent
upper bound of $(D_{x}^B)^2$ and $(D_{y}^A)^2$, respectively.}

%
\section{Quantum set in $D$-space}
\label{sec: Quantum set in D-space}

As mentioned, $C^{A}_x$ is the bias of $p(a|x)$, but it is also the bias of
Bob's optimal probability of guessing Alice's outcome $a$, without the use
of any side information.
In the nonlocality scenario, however, Bob has a half of a shared
state; the local state $\rho_{a|x}$ (conditioned
on Alice's outcome $a$), and by the use of it the guessing
probability is generally increased.
Therefore, it seems another natural way of describing nonlocal correlations
to use the guessing probabilities optimized under 
$\rho_{a|x}$.
For this purpose, we focus on the quantities introduced in
\cite{Ishizaka17a,Ishizaka17aE}
\begin{equation}
D^{B}_x\equiv\max_{\langle X^{2}_B\rangle=1}\langle A_x X_B\rangle,
\hbox{~~}
D^{A}_y\equiv\max_{\langle X^{2}_A\rangle=1}\langle X_A B_y\rangle,
\label{eq: Definition of D}
\end{equation}
where the maximization is taken over any Hermite operator $X_B$ ($X_A$) 
on Bob's (Alice's) side.
Indeed, when $\rho_{1|x}$ and $\rho_{-1|x}$ are both pure
states, the maximum in the definition of $D^{B}_x$ is
attained when $X^{2}_B\!=\!I$ \cite{Ishizaka17a}; hence
$D^{B}_x$ becomes equal to $\hbox{tr}|\rho_{1|x}\!-\!\rho_{-1|x}|$,
coinciding with the bias of Bob's optimal guessing probability \cite{Helstrom69a}.

Let us then describe a correlation by a behavior
$\{\delta^{B}_x,\delta^{A}_y,C_{xy}\}$, such that it is realized by quantum
mechanics if and only if
there exist $|\psi\rangle$, $A^{2}_x\!=\!B^{2}_y\!=\!I$, 
$\delta^{B}_x\!=\!(D^{B}_x)^2$,
$\delta^{A}_y\!=\!(D^{A}_y)^2$, and
$C_{xy}\!=\!\langle \psi |A_x B_y|\psi\rangle$.
The reason for taking the square of $D^{B}_x$ and $D^{A}_y$
will become clear soon.
Such a behavior specifies a point in an 8-dimensional space,
which we denote by the $D$-space.
Note that the behaviors in the $C$-space and the $D$-space have no
one-to-one correspondence. For example, the completely random correlation is
uniquely represented by $\{C^{A}_x\!=\!0,C^{B}_y\!=\!0,C_{xy}\!=\!0\}$
in the $C$-space but represented in the $D$-space by
$\{\delta^{B}_x\!=\!0,\delta^{A}_y\!=\!0,C_{xy}\!=\!0\}$
and
$\{\delta^{B}_x\!=\!1,\delta^{A}_y\!=\!1,C_{xy}\!=\!0\}$.
The former is realized by $A_x\!=\!B_y\!=\!\sigma_1$ on
$|\psi\rangle\!=\!|00\rangle$,
and the latter is realized by $A_x\!=\sigma_1$, $B_y\!=\sigma_3$ on
$|\psi\rangle\!=\!(|00\rangle+|11\rangle)/\sqrt{2}$.

Now, let us investigate the properties of the behaviors in the $D$-space.
When the behaviors $\mathbf{p}_i$ are realized by quantum mechanics, there
always exists a realization of the behavior
$\mathbf{p}\!=\!\sum_i\lambda_i\mathbf{p}_i$
for any $\lambda_i\!\ge\!0$ such that $\sum_i\lambda_i\!=\!1$.
This is because, as shown in Appendix \ref{sec: Convexity of D},
although $(D_{x}^B)^2$ and $(D_{y}^A)^2$ are convex in general such that
\begin{equation}
\left[D_{x}^B(\mathbf{p})\right]^2 \le \sum_i\lambda_i \left[D_{x}^B(\mathbf{p}_i)\right]^2,
\label{eq: Convexity}
\end{equation}
the equality holds, at least when each local state of
the realization of $\mathbf{p}_i$ has orthogonal support, and hence,

{\bf Lemma 1.}
{\it The behaviors $\{\delta^{B}_x,\delta^{A}_y,C_{xy}\}$, which are realized
by quantum mechanics, constitute a convex set.}

This set, denoted by ${\cal Q}_D$, is then at least enclosed by the
hyperplanes in the $D$-space defined from the following inequalities:
\begin{eqnarray}
{\cal B}^{B}&\!\!\!\equiv\!\!\!&-\sum_x V^{B}_x \delta^{B}_x
+\sum_{xy}V^{B}_{xy}C_{xy} \le \frac{1}{4q^{B}},
\label{eq: QBellB} \\
{\cal B}^{A}&\!\!\!\equiv\!\!\!&-\sum_y V^{A}_y \delta^{A}_y
+\sum_{xy}V^{A}_{yx}C_{xy} \le \frac{1}{4q^{A}},
\label{eq: QBellA}
\end{eqnarray}
where the coefficients satisfy
$V^{c}_x\!\ge\!0$, $\prod_{xy}V^{c}_{xy}\!\le\!0$, and
$V^{c}_1V^{c}_{00}V^{c}_{01}\!=\!-V^{c}_0V^{c}_{10}V^{c}_{11}$
for both $c\!=\!A,B$. Note that $V^{B}_{01}$ is the coefficient
of $C_{01}$, but $V^{A}_{01}$ is the coefficient of $C_{10}$.
The quantum bound of the inequalities is given by
\begin{eqnarray}
q^{c}\!=\!\frac{V^{c}_0}{(s^{c}_0)^2}\!=\!\frac{V^{c}_1}{(s^{c}_1)^2},\hbox{~~}
s^{c}_0\!\equiv\!\sqrt{\frac{\lambda^{c}}{V^{c}_{10}V^{c}_{11}}}, \hbox{~~}
s^{c}_1\!\equiv\!\sqrt{\frac{-\lambda^{c}}{V^{c}_{01}V^{c}_{00}}},\nonumber\\
\lambda^{c}\!\equiv\!V^{c}_{10}V^{c}_{11}[(V^{c}_{00})^2\!+\!(V^{c}_{01})^2]
-V^{c}_{00}V^{c}_{01}[(V^{c}_{10})^2\!+\!(V^{c}_{11})^2].
\end{eqnarray}
This is due to the cryptographic quantum bound shown in \cite{Ishizaka17a}.
Indeed, $u^{B}_{xy}\!=\!(-1)^{xy}V^{B}_{xy}/s^{B}_{x}$ 
fulfills $\sum_{xy}(u^{B}_{xy})^2\!=\!1$ and
$u^{B}_{00}u^{B}_{01}\!=\!u^{B}_{10}u^{B}_{11}$; hence any realization
obeys
\begin{eqnarray}
{\cal B}^{B}&\!\!\!=\!\!\!&-q^{B}\sum_x (s^{B}_x D^{B}_x)^2
\!+\!\sum_{xy}s^{B}_{x}u^{B}_{xy}(-1)^{xy}\langle A_xB_y\rangle \cr
&\!\!\!\le\!\!\!&-q^{B}\sum_x (s^{B}_x D^{B}_x)^2
\!+\!\Big[\sum_{x} (s^{B}_x D^{B}_x)^2\Big]^{\frac{1}{2}}
\le \frac{1}{4q^{B}}.
\label{eq: CQB}
\end{eqnarray}
The same holds for ${\cal B}^A$ by using
$u^{A}_{yx}\!=\!(-1)^{xy}V^{A}_{yx}/s^{A}_{y}$.
The inequalities Eqs.\ (\ref{eq: QBellB}) and (\ref{eq: QBellA}) are respected
by any quantum realization, which we denote by {\it quantum} Bell inequalities
in analogy to the Bell inequalities.

It is convenient to introduce another convex set,
which is enclosed by Eqs.\ (\ref{eq: QBellB}) and (\ref{eq: QBellA}).
As Eq.\ (\ref{eq: CQB}) holds whenever the first inequality
due to the cryptographic quantum bound holds,
the behaviors in this set are those satisfying the
TLM inequality Eq.\ (\ref{eq: TLM inequality})
for both scaled correlators
$\tilde C_{xy}\!=\!C_{xy}/\sqrt{\delta^{B}_x}$ and
$\tilde C_{xy}\!=\!C_{xy}/\sqrt{\delta^{A}_y}$ \cite{Ishizaka17a}
[together with the obvious constraint of
$C^{2}_{xy}\!\le\!\delta^{B}_x,\delta^{A}_y\!\le\!1$].
This convex set, denoted by ${\cal Q}_{\rm crypt}$, is a superset of ${\cal Q}_D$.

Let us now search for the extremal points of ${\cal Q}_D$.
It is known that each extremal point of ${\cal Q}_C$ has
a two-qubit realization \cite{Tsirelson80a,Masanes06a}.
This is due to the fact that 
$A_0$ and $A_1$ ($B_0$ and $B_1$ as well) are simultaneously
block-diagonalized by appropriate local bases with the
block size of at most 2 \cite{Masanes06a}.
However, this cannot be applied to the case of ${\cal Q}_D$
due to the convexity of $(D_{x}^B)^2$ and $(D_{y}^A)^2$ as in
Eq.\ (\ref{eq: Convexity}).
Fortunately, however, we have the following:

{\bf Lemma 2.}
{\it A behavior in ${\cal Q}_D$, which simultaneously saturates the
quantum Bell inequalities Eqs.\ (\ref{eq: QBellB}) and (\ref{eq: QBellA}),
has a two-qubit realization.}

\begin{figure}[t]
\centerline{\scalebox{0.55}[0.55]{\includegraphics{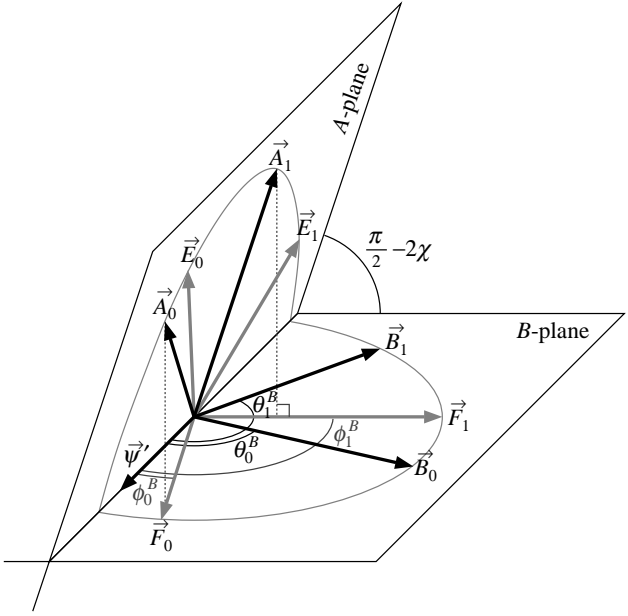}}}
\caption{
Geometry of real vectors, where
$\vec F_0$, $\vec B_0$, $\vec F_1$, and $\vec B_1$
($\vec E_0$, $\vec A_0$, $\vec E_1$, and $\vec A_1$) lie
in the $B$-plane ($A$-plane), and
$\vec F_x$ ($\vec E_y$) is directed along the projection
of $\vec A_x$ ($\vec B_y$) to the $B$-plane ($A$-plane), respectively.
The two planes intersect at the angle of $\pi/2\!-\!2\chi$, with $\vec \psi'$
being a common vector.
The angle between $\vec \psi'$ and $\vec B_x$ ($\vec F_x$) is denoted by
$\theta^{B}_x$ ($\phi^{B}_x$), and that between $\vec\psi'$ and
$\vec A_y$ ($\vec E_y$) is denoted by $\theta^{A}_y$
($\phi^{A}_y$).
As Eq.\ (\ref{eq: TLM inequality}) is saturated by scaled correlators, 
Eq.\ (\ref{eq: TLM condition}) holds in this
geometry.
We assume, without loss of generality, $0\!\le\!\chi\!\le\!\pi/4$
throughout this paper.
}
\label{fig: Geometry}
\end{figure}

{\it Proof:}
As the maximization in $D^{B}_x$ is rewritten by using the Lagrange multiplier
$l$ as
$D^{B}_x\!=\!\max [\langle\psi| A_x X_B|\psi\rangle\!-\!l(\langle\psi| X_B^{2}|\psi\rangle\!-\!1)]$,
any realization must satisfy
\begin{equation}
\hbox{tr}_A A_x|\psi\rangle\langle\psi|
=\frac{D^{B}_x}{2}\hbox{tr}_A (F_x |\psi\rangle\langle\psi|
+|\psi\rangle\langle\psi| F_x),
\label{eq: Extremal condition}
\end{equation}
where $F_x$ is an optimal operator attaining the maximum.
Let $\vec\psi$, $\vec A_x$, $\vec B_y$, and $\vec F_x$ be the real-vector
representation for $|\psi\rangle$, $A_x|\psi\rangle$, $B_y|\psi\rangle$,
and $F_x|\psi\rangle$, respectively, which are all unit vectors.
Then, Eq.\ (\ref{eq: Extremal condition}) implies
\begin{equation}
\vec A_x\cdot \vec F_x \!=\! D^{B}_x, \hbox{~~}
\vec A_x\cdot \vec B_y \!=\! D^{B}_x \vec F_x\cdot\vec B_y, \hbox{~~}
\vec A_x\cdot \vec \psi \!=\! D^{B}_x \vec F_x\cdot\vec \psi.
\label{eq: Inner products}
\end{equation}
On the other hand, the saturation of Eq.\ (\ref{eq: QBellB}) implies that
Eq.\ (\ref{eq: TLM inequality}) is saturated for
$\tilde C_{xy}\!\equiv \!\vec A_x\cdot\vec B_y/D^{B}_x\!=\!\vec F_x\cdot\vec B_y$,
which ensures that four real vectors $\vec B_0$, $\vec B_1$, $\vec F_0$,
and $\vec F_1$ lie in the same $B$-plane \cite{Wang16a} as
shown in Fig.\ \ref{fig: Geometry}.
Similarly, the saturation of Eq.\ (\ref{eq: QBellA}) implies that four real
vectors $\vec A_0$, $\vec A_1$, $\vec E_0$, and $\vec E_1$
lie in the same $A$-plane,
where $\vec E_y$ is the real vector optimizing $D^{A}_y$.
However, as a high-dimensional vector space is considered,
the relationship between the two planes has not been determined yet.

Suppose that $\vec A_0\!\ne\!\pm \vec A_1$ and $\vec B_0\!\ne\!\pm \vec B_1$.
Let the projection of $\vec \psi$ to the $A$-plane ($B$-plane) be
$\vec \psi_{A}$ ($\vec \psi_{B}$).
Moreover, let the projection of $\vec\psi_B$ to the $A$-plane be
$\vec \psi_{BA}$. From the laws of sines and cosines, $|\vec \psi_{BA}|^2$
is given by
\begin{equation}
\frac{(\vec A_0\cdot\vec\psi_B)^2+(\vec A_1\cdot\vec\psi_B)^2
-2(\vec A_0\cdot\vec\psi_B)(\vec A_1\cdot\vec\psi_B)
\cos\Delta}
{\sin^2\Delta},
\end{equation}
where $\Delta$ is the angle between $\vec A_0$ and $\vec A_1$.
From Eq.\ (\ref{eq: Inner products}),
\begin{equation}
\vec A_x\cdot\vec \psi_{B}
=
D^{B}_x \vec F_x\cdot\vec \psi_{B}
=
D^{B}_x \vec F_x\cdot\vec \psi
=
\vec A_x\cdot\vec \psi
=
\vec A_x\cdot\vec \psi_{A},
\end{equation}
and consequently we have $|\vec \psi_{BA}|\!=\!|\vec\psi_{A}|$.
Similarly, we have $|\vec \psi_{AB}|\!=\!|\vec\psi_{B}|$.
This implies that the two planes intersect with
$\vec\psi'\!\equiv\!\vec\psi_A\!=\!\vec\psi_B$ being a common vector
as shown in Fig.\ \ref{fig: Geometry}.
The two-qubit realization of Eq.\ (\ref{eq: Two-qubit realization})
can realize the same geometry of real vectors.

When $\vec A_0\!=\!\pm \vec A_1$, $\vec A_x$ and $\vec B_y$ lie in a
3-dimensional subspace.
Moreover, the saturation of Eq.\ (\ref{eq: TLM inequality}) for scaled
correlators occurs only when $\vec A_0$ coincides with $\pm\vec E_0$
or $\pm\vec E_1$, and $\vec F_0$ coincides with $\pm\vec B_0$
or $\pm\vec B_1$.
The behavior in the $D$-space realized by
such a simple geometry can be realized by
Eq.\ (\ref{eq: Two-qubit realization}).
Similarly, when $\vec B_0\!=\!\pm \vec B_1$.
{\hfill $\Box$}

As such a behavior saturates Eq.\ (\ref{eq: TLM inequality}) for
scaled correlators, it is located at a boundary of ${\cal Q}_{\rm crypt}$.
Conversely, a boundary behavior of ${\cal Q}_{\rm crypt}$ generally does not
have a realization with the geometry of Fig.\ \ref{fig: Geometry} and
cannot be realized by quantum mechanics; hence,

{\bf Lemma 3.}
{\it ${\cal Q}_D$ is a strict subset of ${\cal Q}_{\rm crypt}$.}

Hereafter, to describe the geometry of Fig.\ \ref{fig: Geometry}, we also use
the shortcut notations of
\begin{equation}
\Delta^{c}_{ij}\equiv\phi^{c}_i-\theta^{c}_j,\hbox{~~}
\Delta \theta^{c}\equiv\theta^{c}_0-\theta^{c}_1, \hbox{~~}
\Delta \phi^{c}\equiv\phi^{c}_0-\phi^{c}_1,
\label{eq: Delta}
\end{equation}
for both $c=A,B$.
See Fig.\ \ref{fig: Geometry} for the definition of $\phi^{c}_0$ and $\phi^{c}_1$.
Note that, as Eq.\ (\ref{eq: TLM inequality}) is saturated
for scaled correlators, 
the geometry of Fig.\ \ref{fig: Geometry} satisfies \cite{Ishizaka18a}
\begin{equation}
\prod_{xy}\sin \Delta^{B}_{xy}\le0 \hbox{~~and~~}
\prod_{xy}\sin \Delta^{A}_{yx}\le0.
\label{eq: TLM condition}
\end{equation}
When such a geometry is given,
we can easily construct the quantum Bell inequalities
Eqs.\ (\ref{eq: QBellB}) and (\ref{eq: QBellA}) that are
simultaneously saturated by the geometry, 
as shown in Appendix \ref{sec: Uniqueness of geometry I}.
Conversely, let us investigate the realizations to maximize such a given pair
of the quantum Bell inequalities.
Note that there exists unavoidable ambiguity of the realizations, which is
referred as obvious symmetries hereafter, as
the four geometries with the parameters
$\{\theta^{A}_x,\theta^{B}_y,\chi\}$,
$\{-\theta^{A}_x,-\theta^{B}_y,\chi\}$,
$\{\pi\!-\!\theta^{A}_x,\pi\!-\!\theta^{B}_y,\chi\}$, and
$\{\pi\!+\!\theta^{A}_x,\pi\!+\!\theta^{B}_y,\bar\chi\}$
realize the same behavior in the $D$-space.
In general, the realization that saturates either
Eq.\ (\ref{eq: QBellB}) or Eq.\ (\ref{eq: QBellA}) is not unique; hence
belonging to a flat surface of $Q_D$.
The realization is
characterized by $\Delta \theta^{B}$ and $\Delta \theta^{A}$, respectively,
such that $\Delta^{B}_{xy}$ and $\Delta^{A}_{yx}$ is determined for a given
$\Delta \theta^{B}$ and $\Delta \theta^{A}$, respectively.
However, $\langle A_xB_y\rangle\!=\!D^{B}_x \cos\Delta^{B}_{xy}\!=\!D^{A}_y \cos\Delta^{A}_{yx}$ must hold in Fig.\ \ref{fig: Geometry}.
As a result, to saturates both
Eqs.\ (\ref{eq: QBellB}) and (\ref{eq: QBellA}),
$\Delta \theta^{B}$ and $\Delta \theta^{A}$ are constrained to satisfy
\begin{eqnarray}
\frac{(1+\alpha^{A}\cos\Delta\theta^{A})^2}{(1+\alpha^{A}\cos\Delta\bar\theta^{A})^2}&=&
\frac{(1+\alpha^{B}\cos\Delta\theta^{B})^2}{(1+\alpha^{B}\cos\Delta\bar\theta^{B})^2}, \nonumber \\
\frac{(\cos\Delta\theta^{A}+\alpha^{A})^2}{(\cos\Delta\bar\theta^{A}+\alpha^{A})^2}&=&
\frac{(\cos\Delta\theta^{B}+\alpha^{B})^2}{(\cos\Delta\bar\theta^{B}+\alpha^{B})^2}, \nonumber \\
\frac{(1-\frac{1}{\beta^{A}}\cos\Delta\theta^{A})^2}{(1-\frac{1}{\beta^{A}}\cos\Delta\bar\theta^{A})^2}&=&
\frac{(1-\frac{1}{\beta^{B}}\cos\Delta\theta^{B})^2}{(1-\frac{1}{\beta^{B}}\cos\Delta\bar\theta^{B})^2}, \nonumber \\
\frac{(\cos\Delta\theta^{A}-\frac{1}{\beta^{A}})^2}{(\cos\Delta\bar\theta^{A}-\frac{1}{\beta^{A}})^2}&=&
\frac{(\cos\Delta\theta^{B}-\frac{1}{\beta^{B}})^2}{(\cos\Delta\bar\theta^{B}-\frac{1}{\beta^{B}})^2},
\label{eq: Condition for uniqueness}
\end{eqnarray}
where the parameters of the
original geometry used for constructing a given pair of the Bell
inequalities are indicated by an overline such as $\Delta\bar\theta^{c}$.
The parameters $\alpha^{c}$ and $\beta^{c}$ are given by
$-\sin\bar\Delta^{c}_{00}/\sin\bar\Delta^{c}_{01}$ and
$\sin\bar\Delta^{c}_{11}/\sin\bar\Delta^{c}_{10}$, respectively.
As details are given in Appendix \ref{sec: Uniqueness of geometry I},
when Eq.\ (\ref{eq: Condition for uniqueness}) only has a trivial
solution of $\cos\Delta\theta^{c}\!=\!\cos\Delta\bar\theta^{c}$ for
both $c=A,B$, the realizations become unique up to obvious symmetries,
and we have

{\bf Lemma 4.}
{\it The geometry of a realization, which simultaneously saturates the quantum
Bell inequalities Eqs.\ (\ref{eq: QBellB}) and (\ref{eq: QBellA}),
is unique up to obvious symmetries
when Eq.\ (\ref{eq: Condition for uniqueness})
only has a trivial solution; hence such a behavior is an extremal point of
${\cal Q}_D$.}

For a given pair of quantum Bell inequalities,
no pair of $\alpha^{c}$ and $\beta^{c}$ is identical in general
and Eq.\ (\ref{eq: Condition for uniqueness}) only has a trivial solution.
This implies that the behaviors realized by
two-qubit realizations Eq.\ (\ref{eq: Two-qubit realization})
with the parameters satisfying Eq.\ (\ref{eq: TLM condition})
are generally extremal for ${\cal Q}_D$, constituting a large class of extremal
points.
Note that the uniqueness of the realization is not necessarily
required for the extremality, and hence Lemma 4 does not exclude the
possibility that the behaviors realized by
Eq.\ (\ref{eq: Two-qubit realization}) with Eq.\ (\ref{eq: TLM condition})
are all extremal.

In any case, for an extremal behavior of ${\cal Q}_D$ proven by Lemma 4, the
geometry of real vectors is unique up to the obvious symmetry. Is such a
behavior self-testable?
The answer is negative by two reasons (apart from the
problem of how $D_{x}^B$ and $D_y^{A}$ is determined by experiments).
The first is that $|\vec \psi'|$ in Fig.\ \ref{fig: Geometry}
is undetermined; $|\vec\psi'|$ can be determined through
$\langle A_x\rangle\!=\!|\vec\psi'|\cos\theta^{A}_x$
or $\langle B_y\rangle\!=\!|\vec\psi'|\cos\theta^{B}_y$,
but these are unspecified in the $D$-space.
The second relates to the convexity of $(D_{x}^B)^2$ and $(D_{y}^A)^2$.
As shown in Appendix \ref{sec: Example of strict convexity},
there exists an example in which the correlation $\mathbf{P}$,
despite being an extremal point of ${\cal Q}_D$,
may have two different realizations due to the strict convexity.
However, in some cases, we can exclude the possibility of such strict
convexity, that is, the certifiability of $D_{x}^B$ and $D_{y}^A$.

Suppose that Conjecture 2 holds true.
As Eq.\ (\ref{eq: TLM inequality}) is saturated for scaled
correlators, $(D_{x}^B)^2$ and $(D_{y}^A)^2$ are also lower bounded by
$d^{B}_x$ and $d^{A}_y$ \cite{Ishizaka18a}; hence 
those are certifiable,
and we have $(D^{B}_{x})^2\!=\!d^{B}_x$ and $(D^{A}_{y})^2\!=\!d^{A}_y$.
This correlation, denoted by $\mathbf{p}$, is then found to be an extremal
point of ${\cal Q}_D$ by Lemma 4.
When a realization of $\mathbf{p}$ is decomposed into two-qubit realizations
of $\mathbf{p}_i$, based on the block-diagonalization \cite{Masanes06a},
$(D_{x}^B)^2$ and $(D_{y}^A)^2$ must not be strictly convex;
otherwise we would construct a realization whose 
$D_{x}^B$ or $D_{y}^A$ exceeds the
device-independent upper bound by using orthogonal bases.
Moreover, because the correlation $\mathbf{p}$ is an extremal point of
${\cal Q}_D$, all $\mathbf{p}_i$ must exhibit the same behavior
$\{d^{B}_x,d^{A}_y,C_{xy}\}$ in the $D$-space.
Then, the geometry of the {\it two-qubit} realizations is uniquely determined 
up to the obvious symmetry by Lemma 4.
The symmetry leaves the ambiguity between
$\{C^{A}_x,C^{B}_y\}$ and $\{-C^{A}_x,-C^{B}_y\}$, but the latter is
clearly inappropriate.
In this way, the extremality of ${\cal Q}_D$, combined with the
certifiability of $D_{x}^B$ and $D_{y}^A$, makes the realization unique; hence,

{\bf Lemma 5.}
{\it If Conjecture 2 holds true, the extremal behaviors of ${\cal Q}_D$ by
Lemma 4 are self-testable extremal points of ${\cal Q}_C$.
}

This lemma implies that the sufficiency of Conjecture 1 relies on
the validity of Conjecture 2.
Note that, under the truth of Conjecture 2,
the self-testable extremal points of ${\cal Q}_C$ by Lemma 5
are such that
Eq.\ (\ref{eq: Two-qubit condition}) is satisfied
as $S^{+}_{00}\!=\!S^{+}_{01}\!=\!S^{+}_{10}\!=\!S^{+}_{11}$,
Eq.\ (\ref{eq: TLM inequality}) is saturated by both
$\tilde C_{xy}\!=\!C_{xy}/\sqrt{d^{B}_{x}}$
and $\tilde C_{xy}\!=\!C_{xy}/\sqrt{d^{A}_{y}}$,
and Eq.\ (\ref{eq: Condition for uniqueness}) only has a trivial solution.
As mentioned above,
the information of
$\{C_{x}^A,C_{y}^B\}$ is necessary for self-testing to specify $|\psi'|$,
and it is indeed used in Lemma 5 through
Eq.\ (\ref{eq: Two-qubit condition}).

%
\section{Quantum set in $C$-space}
\label{sec: Quantum set in C-space}

From now on, let us show some geometrical properties of the realizations for
the behaviors in the standard $C$-space. Note that these hold true regardless
of the validity of Conjectures 1 and 2. 
To begin with, we show that the geometry of the realization of a behavior in
the $C$-space is uniquely determined when the correlators satisfy a condition:

{\bf Lemma 6.}
{\it For a nonlocal behavior $\{C^{A}_x,C^{B}_y,C_{xy}\}$,
which satisfies Eq.\ (\ref{eq: Two-qubit condition}) for some $\{p_{xy}\}$
(not necessarily as $S^{+}_{00}\!=\!S^{+}_{01}\!=\!S^{+}_{10}\!=\!S^{+}_{11}$)
and saturates
Eq.\ (\ref{eq: TLM inequality})
for both $\tilde C_{xy}\!=\!C_{xy}/\sqrt{d^{B}_{x}}$
and  $\tilde C_{xy}\!=\!C_{xy}/\sqrt{d^{A}_{y}}$,
the geometry of the realization is unique up to obvious
symmetries.}

The unique geometry is the same as Fig.\ \ref{fig: Geometry},
but the obvious symmetry now refers the ambiguity between
$\{\theta^{A}_x,\theta^{B}_y,\chi\}$ and
$\{-\theta^{A}_x,-\theta^{B}_y,\chi\}$.
Moreover, $|\psi'|$ is determined to $\cos2\chi$ as in the two-qubit
realizations of Eq.\ (\ref{eq: Two-qubit realization}).
The proof is given in Appendix \ref{sec: Uniqueness of geometry II}.
The difference from the proof of Lemma 2 is that $d^{B}_x$ and
$d^{A}_y$ by Eq.\ (\ref{eq: d quantity}) are not ensured to coincide with
$(D^{B}_x)^2$ and $(D^{A}_y)^2$,
and we cannot use Eq.\ (\ref{eq: Extremal condition}).
For the same reason, $\vec F_x$ and $\vec E_y$ in Fig.\ \ref{fig: Geometry}
are now not ensured to attain $D^{B}_x$ and $D^{A}_y$;
$\sqrt{d^{B}_x}\vec F_x$ is merely the projection
of $\vec A_x$ to the $B$-plane.

In this way, the geometry is uniquely determined for not necessarily
$S^{+}_{00}\!=\!S^{+}_{01}\!=\!S^{+}_{10}\!=\!S^{+}_{11}$.
However, this uniqueness does not ensure the
extremality of ${\cal Q}_C$. This is in contrast to Lemma 4, where quantum Bell
inequalities are maximized by a unique geometry, and the extremality of
${\cal Q}_D$ is ensured.
Indeed, the nonlocal correlation $\mathbf{P}$ in
Appendix \ref{sec: Example of strict convexity}, where $S^{+}_{00}\!=\!S^{+}_{01}\!=\!S^{+}_{10}\!=\!S^{-}_{11}$, is an explicit counter example for
extremality.
Interestingly, $\mathbf{P}$ is located in the strict interior of the
quantum set, according to the $1\!+\!AB$ level of the NPA
hierarchy \cite{Ozeki19a}.
This also implies that,
even though $|\psi'|\!=\cos2\chi$ is ensured to be the same
as the two-qubit realizations,
the uniqueness is still insufficient for self-testing.
The condition $S^{+}_{00}\!=\!S^{+}_{01}\!=\!S^{+}_{10}\!=\!S^{+}_{11}$ 
is crucial, apart from the unique determination of the geometry,
for making the realization self-testable through the certification of 
$D_{x}^B$ and $D_{y}^A$, as shown by Lemma 5.

However, other than the unproved certification condition, a more general
condition that makes the unique geometry self-testable is found as follows:

{\bf Lemma 7.}
{\it For a nonlocal behavior $\{C^{A}_x,C^{B}_y,C_{xy}\}$,
which has a unique geometry by Lemma 6,
the realization is
self-testable, if and only if 
a real vector representation $\vec G$ of $G|\psi\rangle$, with $G$ being a
local unitary observable, exists in either $A$-plane or $B$-plane
(other than $\pm\vec A_x$ and $\pm\vec B_y$).
}

\begin{figure}[t]
\centerline{\scalebox{0.55}[0.55]{\includegraphics{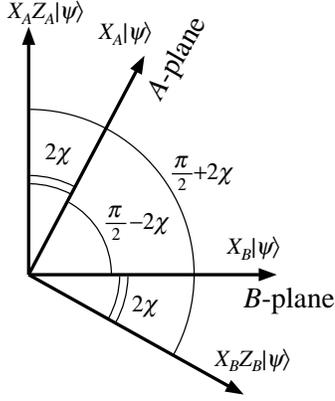}}}
\caption{
Four state vectors lie in the same plane, and
$X_B|\psi\rangle$ and $X_AZ_A|\psi\rangle$ are orthogonal,
which also implies that $Z_A|\psi\rangle$ and $X_AX_B|\psi\rangle$
are orthogonal because $\langle\psi|X_BX_AZ_A|\psi\rangle\!=\!0$.
}
\label{fig: Real vectors 2}
\end{figure}

\begin{figure}[t]
\centerline{\scalebox{0.5}[0.5]{\includegraphics{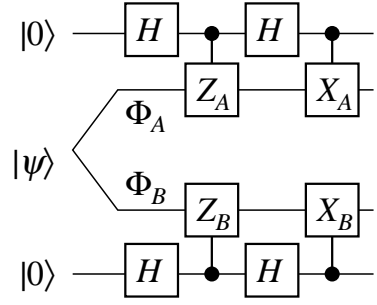}}}
\caption{
Local unitary transformation used for self-testing.
}
\label{fig: selftesting}
\end{figure}

{\it Proof:}
As the geometry is uniquely determined as Fig.\ \ref{fig: Geometry} by
Lemma 6, the ``only if'' part is obvious:
when the realization is self-testable,
it is a two-qubit realization of Eq.\ (\ref{eq: Two-qubit realization}),
where any one of $\vec F_x$ and $\vec E_y$ can be regarded as $\vec G$
because $F_x$ and $E_y$ are local and unitary ($F_x^{2}\!=\!E_y^{2}\!=\!I$).
Let us prove the ``if'' part.
We again use the notation of Eq.\ (\ref{eq: Delta}).
For the operator $Z_B$ defined by
\begin{equation}
Z_B=\frac{
\sin\theta^{B}_0B_1-\sin\theta^{B}_1B_0}
{\sin\Delta \theta^{B}},
\label{eq: Z_B}
\end{equation}
we have $\langle\psi|Z^{2}_B|\psi\rangle\!=\!1$ as $\vec B_0\cdot\vec B_1\!=\!\cos\Delta\theta^B$, and similarly for $Z_A$.
As the unit vectors $Z_B|\psi\rangle$ and $Z_A|\psi\rangle$ are both directed along $\vec\psi'$,
we have $Z_B|\psi\rangle\!=\!\frac{1}{\cos2\chi}|\psi'\rangle\!=\!Z_A|\psi\rangle$.
Suppose now that $\vec G$ lies in the $B$-plane
with $G$ being Bob's unitary observable ($G^{2}\!=\!I$).
Letting the angle between $\vec G$ and $\vec \psi'$ be $\eta^B$, it is
written as
\begin{equation}
G|\psi\rangle=\frac{
\sin\eta^B B_y|\psi\rangle-\sin(\eta^B-\theta_{y}^{B})Z_B|\psi\rangle}
{\sin \theta^{B}_y},
\end{equation}
for $y\!=\!0,1$.
Moreover, as $G$ commutes with $Z_A$ and $G^{2}\!=\!I$, we have
$\langle\psi|GZ_AZ_AG|\psi\rangle\!=\!\langle\psi|Z^{2}_A|\psi\rangle
\!=\!1$ and
\begin{eqnarray}
\sin^2\theta^{B}_y&\!\!\!=\!\!\!&
\sin^2\eta^{B}+\sin^2(\eta^B-\theta_{y}^B)\langle\psi|Z^{4}_B|\psi\rangle \cr
&\!\!\!\!\!\!&
-2\sin\eta^{B}\sin(\eta^B-\theta_{y}^B)\langle\psi|Z^{3}_B B_y|\psi\rangle.
\end{eqnarray}
From this and Eq.\ (\ref{eq: Z_B}), we have
$\langle\psi|Z^{4}_B|\psi\rangle\!=\!1$; hence $Z^{2}_B|\psi\rangle$ is a unit
vector.
As $\langle\psi|Z^{2}_B|\psi\rangle\!=\!1$,
we have $Z^{2}_B|\psi\rangle\!=\!|\psi\rangle$, which
proves the anti-commutation relation of
\begin{equation}
(B_0B_1+B_1B_0)|\psi\rangle=2\cos\Delta\theta^{B}|\psi\rangle.
\label{eq: Anticommutation}
\end{equation}
As $Z^{2}_A|\psi\rangle\!=\!Z^{2}_B|\psi\rangle\!=\!|\psi\rangle$,
the anti-commutation relation between $A_0$ and $A_1$ is also proven.
Let us define $X_B$ by
\begin{equation}
X_B=\frac{
\cos\theta^{B}_0B_1-\cos\theta^{B}_1B_0}
{\sin\Delta \theta^{B}}.
\end{equation}
and similarly $X_A$.
With the anti-commutation relations of $B_y$ and $A_x$,
we can confirm
$(X_B)^{2}|\psi\rangle\!=\!(X_A)^{2}|\psi\rangle\!=\!|\psi\rangle$ and
$(X_BZ_B\!+\!Z_BX_B)|\psi\rangle\!=\!(X_AZ_A\!+\!Z_AX_A)|\psi\rangle\!=\!0$.
However, $|\psi\rangle$ has not been determined yet.
From Eq.\ (\ref{eq: C_xy2}), we have
\begin{eqnarray}
\langle\psi|Z_AX_AX_BZ_B|\psi\rangle
&\!\!\!=\!\!\!&
-\langle\psi|Z_AX_AZ_BX_B|\psi\rangle \cr
&\!\!\!=\!\!\!&-\langle\psi|X_AX_B|\psi\rangle=-\sin2\chi, \cr
\langle\psi|X_AX_AZ_A|\psi\rangle&\!\!\!=\!\!\!&\langle\psi|X_BX_BZ_B|\psi\rangle=\cos2\chi.
\end{eqnarray}
This implies that the four state vectors
(not in the real-vector representation)
of
$X_AZ_A|\psi\rangle$,
$X_A|\psi\rangle$,
$X_B|\psi\rangle$,
and $X_BZ_B|\psi\rangle$ lie in the same plane in a complex vector space,
as shown in Fig.\ \ref{fig: Real vectors 2}.
Moreover, this figure shows that $\langle\psi|X_BX_AZ_A|\psi\rangle\!=\!0$;
hence $Z_A|\psi\rangle$ and $X_AX_B|\psi\rangle$ are orthogonal to each
other.
As the components of $|\psi\rangle$ to these orthogonal vectors are
given by $\langle\psi|Z_A|\psi\rangle\!=\!\cos2\chi$ and
$\langle\psi|X_AX_B|\psi\rangle\!=\!\sin\chi$, we can conclude
\begin{equation}
|\psi\rangle=\sin2\chi X_AX_B|\psi\rangle+\cos2\chi Z_A|\psi\rangle.
\end{equation}
By operating $X_AX_B$ on both sides, we have
\begin{eqnarray}
\sin2\chi X_AX_BZ_A|\psi\rangle&\!\!\!=\!\!\!&
\frac{\sin2\chi X_AX_B|\psi\rangle-\sin^22\chi|\psi\rangle}{\cos2\chi}
\nonumber\\
&\!\!\!=\!\!\!&\cos2\chi |\psi\rangle-Z_A|\psi\rangle,
\end{eqnarray}
and
$\cos\chi X_AX_B(I\!-\!Z_A)|\psi\rangle\!=\!\sin\chi (I\!+\!Z_A)|\psi\rangle$.
Then, the local unitary transformation $\Phi\!\equiv\!\Phi_A\otimes\Phi_B$
commonly used for self-testing \cite{Scarani12a} shown in
Fig.\ \ref{fig: selftesting} results in
\begin{eqnarray}
\Phi|\psi\rangle|00\rangle&\!\!\!=\!\!\!&
\frac{1}{4}\big[(I+Z_A)(I+Z_B)|\psi\rangle|00\rangle \cr
&&\quad+X_B(I+Z_A)(I-Z_B)|\psi\rangle|01\rangle \cr
&&\quad+X_A(I-Z_A)(I+Z_B)|\psi\rangle|10\rangle \cr
&&\quad+X_AX_B(I-Z_A)(I-Z_B)|\psi\rangle|11\rangle\big] \cr
&\!\!\!=\!\!\!&\frac{(I+Z_A)|\psi\rangle}{2\cos\chi}
(\cos\chi|00\rangle+\sin\chi|11\rangle),
\end{eqnarray}
and consequently $|\psi\rangle$ is locally equivalent to 
$\cos\chi|00\rangle\!+\!\sin\chi|11\rangle$. Similarly, we also have
\begin{equation}
\Phi X_A X_B|\psi\rangle|00\rangle=|\hbox{junk}\rangle
(\cos\chi|11\rangle+\sin\chi|00\rangle),
\end{equation}
and so on, and measurements are self-tested. {\hfill $\Box$}

For self-testability, the proof of the anti-commutation relation
between $B_0$ and $B_1$ [Eq.\ (\ref{eq: Anticommutation})] is crucial.
To prove it, Lemma 7 implies that the third unitary observable $G$, whose real
vector lies in the same $B$-plane, is necessary.
In the unbiased case where $\chi\!=\!\pi/4$, the four vectors
$\vec A_0$, $\vec A_1$, $\vec B_0$,
$\vec B_1$ all lie in the same plane,
and $A_x$ can be used as the third unitary observable \cite{Wang16a}.
However, in the other general case of $0\!<\!\chi\!<\!\pi/4$,
$\vec A_0$ and $\vec A_1$ lie in a different $A$-plane, and $A_x$ cannot be
used anymore.
It is not limited, but the optimal operator $F_x$ for $D^{B}_x$ is a good
candidate for $G$. Interestingly, in the special case that
$\vec F_0\!=\!\vec B_0$ and $\vec F_1\!=\!\vec B_1$, the candidate for $G$
is missing in the $B$-plane, but the correlation in this case
is always local.


%
\section{Scheme for self-testing partially entangled state}
\label{sec: Scheme for self-testing partially entangled state}

As shown in Sec.\ \ref{sec: Quantum set in D-space}, under the conjectured
certifiability of $D_{x}^B$ and $D_{y}^A$, the realizations are automatically
self-testable by Lemma 5; however, Conjecture 2 has not been proven.
Fortunately, however, Lemma 7 tells us how to self-test
such realizations irrespective of the validity of the conjecture; it suffices
to intentionally  introduce a unitary observable by adding one more binary
measurement.

The simplest protocol may be to add the measurement of $Z_B$.
Let us add a binary measurement to the Bell scenario, such as the Bell
$(2,3,2)$-scenario but on Bob's side only, whose observable is $B_2$
($B^{2}_2\!=\!I$).
Suppose that the correlators by the original set $\{A_0,A_1,B_0,B_1\}$
satisfy the condition in Lemma 6, and the geometry of real vectors is
determined as Fig.\ \ref{fig: Geometry}, where $\sin2\chi$ is also determined.
When the additional correlators satisfy
\begin{equation}
\langle A_x B_2\rangle=\cos\theta^{A}_x=\langle A_x\rangle/\cos2\chi,\hbox{~~}
\langle B_2 \rangle=\cos2\chi,
\end{equation}
for both $x\!=\!0,1$, $\vec B_2$ is ensured to lie in the $A$-plane and is
 directed along $\vec \psi'$.
Then, in this protocol, 
$B_2|\psi\rangle\!=\!Z_B|\psi\rangle\!=\!Z_A|\psi\rangle$ can be directly used
for proving the anti-commutation relation of $B_y$ ($A_x$ also)
as in the proof of Lemma 7.

The additional measurement is not restricted to $Z_B$.
In the second protocol, suppose that the correlators by $\{A_0,A_1,B_0,B_2\}$
also satisfy the condition in Lemma 6,
in addition to the original $\{A_0,A_1,B_0,B_1\}$.
Then, as $\vec \psi'$, $\vec B_0$, $\vec B_2$ lie in the same plane,
$\vec B_2$ is ensured to lie in the $B$-plane, and again, $B_2$ can be
used as the third observable for proving the anti-commutation relation
between $B_0$ and $B_1$; the proof of Lemma 7 runs similarly, and the
realization is self-tested.

Note that $B_2$ is also self-tested at the end of both
protocols.
Obviously, the scheme of the second protocol can be repeated to add more
measurements on both sides of Alice and Bob.
In this way, by using the geometry of Fig.\ \ref{fig: Geometry}
as a building block,
the two-qubit realizations in the form of
Eq.\ (\ref{eq: Two-qubit realization})
with arbitrary number of measurements (whose basis lies in the
$X$-$Z$ plane) can be self-tested.

%
\section{Summary}
\label{sec: Summary}

In this paper, we studied the self-testability and extremality from the
viewpoint of the geometry of the state vectors of the realizations
for quantum correlations, and showed a condition that determines the
geometry uniquely.
Interestingly, in the case of the realizations using partially entangled
states, the condition for the unique determination of the geometry is strictly
looser than that for the self-testability.

We first showed that the saturation of the TLM inequality for scaled
correlators, together with the existence of a two-qubit realization in the
form of Eq.\ (\ref{eq: Two-qubit realization}), uniquely determines the
geometry of state vectors in both cases of the $D$-space and
the $C$-space (Lemma 4 and Lemma 6).
The uniqueness of the geometry generally ensures the extremality of
${\cal Q}_D$,
because it is a unique simultaneous maximizer of two quantum Bell inequalities
in the $D$-space.
In the case of the $C$-space, however, such quantum Bell inequalities are
lacking, and the uniqueness of the geometry is insufficient for the extremality
of ${\cal Q}_C$.
Indeed, there exists a two-qubit realization such that,
despite being an extremal point of ${\cal Q}_D$,
it is not an extremal point of ${\cal Q}_C$ due to the convexity of guessing
probabilities.
This suggests that the structure of ${\cal Q}_D$
is simpler than ${\cal Q}_C$.
The complete characterization of the extremal points of ${\cal Q}_D$ 
is an intriguing open problem.

We next showed that, when the conjectured certifiability of the guessing
probabilities holds true,
the self-testability in the $C$-space
(hence the extremality of ${\cal Q}_C$) comes to be ensured by the
extremality of ${\cal Q}_D$ (Lemma 5).
Namely, the sufficiency of the extremality criterion conjectured in
\cite{Ishizaka18a} was shown to rely on the certifiability of
guessing probabilities.
The proof of the certifiability  (i.e., the proof of the device-independent
upper bound of guessing probabilities) seems quite challenging but attractive,
because it would also lead to the discovery of the information
principles \cite{Brunner14a,Chiribella16a}
behind quantum mechanics, and ``almost quantumness'' \cite{Navascues15a}
as well.

Moreover, the realization with a unique geometry becomes self-testable if and
only if another unitary observable exists on the geometry (Lemma 7).
Applying this proven fact,
we proposed self-testing protocols for partially entangled two-qubit
states, where one more measurement
is intentionally added to prove the anti-commutation relation between
observables.
This geometrical scheme provides a building block used for a
more complicated geometry.
Indeed, repeatedly adding measurements by this scheme, a realization with
an arbitrary number of measurements can be self-tested.
It is an open problem of how robust this scheme is.

As all the known nonlocal extremal points in the simplest Bell scenario
are self-testable, it is natural to expect that the true extremal criterion
must be the one that determines the geometry of state vectors as well as
the TLM criterion.
The conjectured criterion in \cite{Ishizaka18a} fulfills this
expectation. Interestingly, although the validity of the conjecture has
not been proven, the property of determining the geometry
proves the self-testability of the realizations in the
Bell scenario with more measurement settings as in the above 
self-testing protocols.

\begin{acknowledgments}
This work was supported by JSPS KAKENHI Grant No. 17K05579. 
\end{acknowledgments}

%
\appendix
%
\section{Convexity of $D_{x}^B$}
\label{sec: Convexity of D}
Let $\rho_{a|x}$ be Bob's subnormalized conditional state.
For any convex decomposition
$\rho_{a|x}\!=\!\sum_i \lambda_i\rho^{(i)}_{a|x}$, we have
\begin{eqnarray}
D^{B}_x&\!\!\!=\!\!\!&
\max_{X_B}\hbox{tr}(\rho_{1|x}\!-\!\rho_{-1|x})X_B  \cr
&\!\!\!=\!\!\!&\max_{X_B}\sum_i\lambda_i
\hbox{tr}(\rho^{(i)}_{1|x}\!-\!\rho^{(i)}_{-1|x})X_B \cr
&\!\!\!=\!\!\!&\max_{X_B}\sum_i\lambda_i
\hbox{tr}[(\rho^{(i)}_{1|x}\!-\!\rho^{(i)}_{-1|x})\!\otimes\!(|i\rangle\langle i|)_a](X_B\!\otimes\!I_a) \cr
&\!\!\!\le\!\!\!&\max_{X_{Ba}}\sum_i\lambda_i
\hbox{tr}[(\rho^{(i)}_{1|x}\!-\!\rho^{(i)}_{-1|x})\otimes (|i\rangle\langle i|)_a]X_{Ba} \cr
&\!\!\!=\!\!\!&\bigg[\sum_i\lambda_i \Big[\max_{X^{(i)}_B}
\hbox{tr}(\rho^{(i)}_{1|x}\!-\!\rho^{(i)}_{-1|x})X^{(i)}_B\Big]^2\bigg]^{1/2},
\end{eqnarray}
where $a$ denotes the ancilla. At the last equality,
we used the formula
$(D_{x}^B)^2\!=\!
\sum_{kk'}2|a_{kk'}|^2/(m_k+m_{k'})$,
where $a_{kk'}\!=\!\langle k|(\rho_{1|x}\!-\!\rho_{-1|x})|k'\rangle$ are the
matrix elements with respect to the eigenstates of
$\rho_{1|x}\!+\!\rho_{-1|x}$ with $m_k$ and $m_{k'}$ being the eigenvalues, as
shown in Appendix A of \cite{Ishizaka17a}. See also \cite{Ishizaka17aE}.

%
\section{Uniqueness of geometry I}
\label{sec: Uniqueness of geometry I}

First, we explicitly show how to construct a pair of the quantum Bell
inequalities Eqs.\ (\ref{eq: QBellB}) and (\ref{eq: QBellA}) that is
simultaneously saturated by a given geometry of Fig.\ \ref{fig: Geometry}
(i.e. a given set of the geometrical parameters
$\{\theta^{A}_x,\theta^{B}_y,\chi\}$).
The saturation condition for the first inequality in
Eq.\ (\ref{eq: CQB}) is that,
for $X_x\equiv\sum_y u^{B}_{xy}(-1)^{xy}B_y$,
\begin{eqnarray}
&& X_x \propto F_x=\frac{(\sin\Delta^{B}_{x1}B_0-\sin\Delta^{B}_{x0}B_1)}
{\sin\Delta\theta^{B}}, \nonumber \\
&&(s^{B}_0 D^{B}_0)^2\langle X^{2}_1\rangle=
(s^{B}_1 D^{B}_1)^2\langle X^{2}_0\rangle,
\end{eqnarray}
and the coefficients of the quantum Bell inequalities must satisfy
\begin{eqnarray}
&\!\!\!\!\!\!&u^{c}_{00}\sin\Delta^{c}_{00}=-u^{c}_{01}\sin\Delta^{c}_{01}, \hbox{~~}
u^{c}_{10}\sin\Delta^{c}_{10}=u^{c}_{11}\sin\Delta^{c}_{11}, \nonumber \\
&\!\!\!\!\!\!&(s^{c}_0 D^{c}_0)^2|\sin\Delta^{c}_{01}\sin\Delta^{c}_{00}|
=(s^{c}_1 D^{c}_1)^2|\sin\Delta^{c}_{11}\sin\Delta^{c}_{10}|,  \nonumber \\
&\!\!\!\!\!\!&(s^{c}_0 D^{c}_0)^2+(s^{c}_1 D^{c}_1)^2 =\frac{1}{4(q^{c})^2},
\label{eq: Unique condition}
\end{eqnarray}
where the last equation is the saturation condition for the second
inequality of Eq.\ (\ref{eq: CQB}). It is then sufficient to choose for both
$c=A,B$ as follows:
\begin{eqnarray}
&\!\!\!\!\!\!&u^{c}_{00}=a\sin\Delta^{c}_{01},\hbox{~~} u^{c}_{01}=-a\sin\Delta^{c}_{00}, \nonumber \\
&\!\!\!\!\!\!&u^{c}_{10}=b\sin\Delta^{c}_{11},\hbox{~~} u^{c}_{11}=b\sin\Delta^{c}_{10}, \nonumber \\
&\!\!\!\!\!\!& s^{c}_0= D^{c}_1 a,\hbox{~~}s^{c}_1= D^{c}_0 b, \hbox{~~}
1/q^{c}=2\sqrt{(s^{c}_0 D^{c}_0)^2+(s^{c}_1 D^{c}_1)^2}, \nonumber \\
&\!\!\!\!\!\!&a=\frac{1}{\sin\Delta\theta^{c}}\sqrt{
\frac{\sin\Delta^{c}_{11}\sin\Delta^{c}_{10}}
{\sin\Delta^{c}_{11}\sin\Delta^{c}_{10}-\sin\Delta^{c}_{01}\sin\Delta^{c}_{00}}
}, \nonumber \\
&\!\!\!\!\!\!&b=\frac{1}{\sin\Delta\theta^{c}}\sqrt{
\frac{-\sin\Delta^{c}_{01}\sin\Delta^{c}_{00}}
{\sin\Delta^{c}_{11}\sin\Delta^{c}_{10}-\sin\Delta^{c}_{01}\sin\Delta^{c}_{00}}
}.
\end{eqnarray}

Next, let us show conversely that, for a given set of such
coefficients of quantum Bell inequalities,
the geometrical
parameters satisfying Eq.\ (\ref{eq: Unique condition}) are unique
(up to obvious symmetries). Let
$\alpha^{c}\!\equiv\!u^{c}_{01}/u^{c}_{00}$
and $\beta^{c}\!\equiv\!u^{c}_{10}/u^{c}_{11}$.
Once we choose $\Delta\theta^{c}$,
$\tan\Delta^{c}_{ij}$ is determined from Eq.\ (\ref{eq: Unique condition})
as
\begin{eqnarray}
&&\tan\Delta^{c}_{00}=\frac{-\alpha^{c}\sin\Delta\theta^{c}}{1+\alpha^{c}\cos\Delta\theta^{c}},\hbox{~~}
\tan\Delta^{c}_{01}=\frac{\sin\Delta\theta^{c}}{\cos\Delta\theta^{c}+\alpha^{c}},\nonumber \\
&&\tan\Delta^{c}_{10}=\frac{\frac{1}{\beta^{c}}\sin\Delta\theta^{c}}{1-\frac{1}{\beta^{c}}\cos\Delta\theta^{c}}, \hbox{~~}
\tan\Delta^{c}_{11}=\frac{\sin\Delta\theta^{c}}{\cos\Delta\theta^{c}-\frac{1}{\beta^{c}}}, \nonumber
\end{eqnarray}
and as a result, $D_0^{c}$  and $D_1^{c}$ is also determined by
$\Delta\theta^c$ as
\begin{eqnarray}
(D^{c}_0)^2&\!\!\!=\!\!\!&
\frac{1}{4(s^{c}_0q^{c})^2}
\frac{\alpha^{c}+\frac{1}{\alpha^{c}}+2\cos\Delta\theta^{c}}
{\alpha^{c}+\frac{1}{\alpha^{c}}+\beta^{c}+\frac{1}{\beta^{c}}}, \nonumber \\
(D^{c}_1)^2&\!\!\!=\!\!\!&
\frac{1}{4(s^{c}_1q^{c})^2}
\frac{\beta^{c}+\frac{1}{\beta^{c}}-2\cos\Delta\theta^{c}}
{\alpha^{c}+\frac{1}{\alpha^{c}}+\beta^{c}+\frac{1}{\beta^{c}}}.
\label{eq: determine D}
\end{eqnarray}
For these solutions to represent the same realization,
$\langle A_xB_y\rangle^2\!=\!(D^{B}_x \cos\Delta^{B}_{xy})^2\!=\!(D^{A}_y \cos\Delta^{A}_{yx})^2$
must hold for every $x$ and $y$;
hence Eq.\ (\ref{eq: Condition for uniqueness}) must hold,
where the original geometrical parameters appears in
Eq.\ (\ref{eq: Unique condition})
are indicated by an overline.
When Eq.\ (\ref{eq: Condition for uniqueness}) only has a trivial solution
of $\cos\Delta\theta^{c}\!=\!\cos\Delta\bar\theta^{c}$,
we have $D^{c}_0\!=\!\bar D^{c}_0$ and $D^{c}_1\!=\!\bar D^{c}_1$
from Eq.\ (\ref{eq: determine D}).
Moreover, from $D^{B}_0D^{B}_1\sin\Delta\phi^{B}\!=\!\sin2\chi\sin\Delta\theta^{A}$ and $0\!\le\!\chi\!\le\!\pi/4$,
we have $\chi\!=\!\bar\chi$ as 
$\tan\Delta\phi^{c}\!=\!\tan(\Delta^{c}_{00}\!-\!\Delta^{c}_{10})
\!=\!\pm\tan\Delta\bar\phi^{c}$.
From $(D^{B}_x)^2\!=\!\cos^22\chi\cos^2\theta^{A}_y\!+\!\sin^22\chi$,
we have $\cos^2\theta^{A}_y\!=\!\cos^2\bar\theta^{A}_y$,
and similarly $\cos^2\theta^{B}_x\!=\!\cos^2\bar\theta^{B}_x$.
Considering the possible combination of signs carefully, it is found that
the allowed solutions of Eq.\ (\ref{eq: Unique condition}) are only
$\{\bar\theta^{A}_x,\bar\theta^{B}_y,\bar\chi\}$,
$\{-\bar\theta^{A}_x,-\bar\theta^{B}_y,\bar\chi\}$,
$\{\pi\!-\!\bar\theta^{A}_x,\pi\!-\!\bar\theta^{B}_y,\bar\chi\}$, and
$\{\pi\!+\!\bar\theta^{A}_x,\pi\!+\!\bar\theta^{B}_y,\bar\chi\}$.

%
\section{Example of strict convexity}
\label{sec: Example of strict convexity}

\begin{figure}[t]
\centerline{\scalebox{0.55}[0.55]{\includegraphics{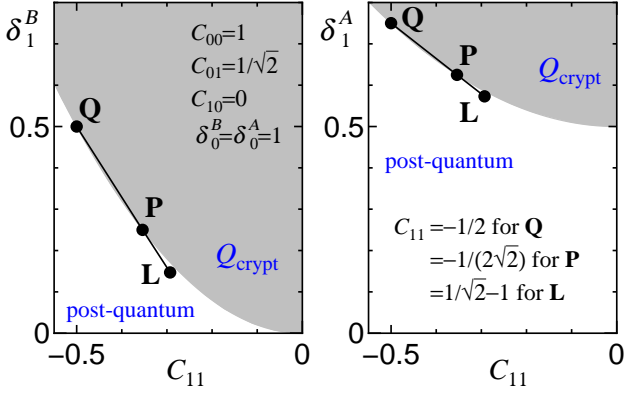}}}
\caption{
Two cross sections of the $D$-space, $C_{11}$-$\delta^B_1$ and
$C_{11}$-$\delta^A_1$,
which contain $\mathbf{P}$, $\mathbf{Q}$, and $\mathbf{L}$.
The behaviors in the gray region belong to ${\cal Q}_{\rm crypt}$.
When $\mathbf{P}\!=\!\lambda \mathbf{Q}\!+\!(1\!-\!\lambda)\mathbf{L}$
with $\lambda\!=\!1\!-\!\frac{1}{\sqrt{2}}$
(in the limit of $\epsilon\!\rightarrow\!0$), the behavior of $\mathbf{L}$ in
the $D$-space is not quantum realizable, whereas the behavior in the $C$-space
is locally realizable.
}
\label{fig: PQL}
\end{figure}

Let us consider the two nonlocal correlations
$\mathbf{P}$ and $\mathbf{Q}$ realized by
Eq.\ (\ref{eq: Two-qubit realization}) using the following parameters:
\begin{eqnarray*}
&\!\!\!\!\!\!& 
\mathbf{P}\hbox{:~} \theta^{A}_0\!=\!0,\hbox{~} \theta^{A}_1\!=\!\pi/2,\hbox{~} \theta^{B}_0\!=\!\epsilon,\hbox{~} \theta^{B}_1\!=\!-\pi/4,\hbox{~} 2\chi\!=\!\pi/6, \\
&\!\!\!\!\!\!& 
\mathbf{Q}\hbox{:~} \theta^{A}_0\!=\!0,\hbox{~} \theta^{A}_1\!=\!\pi/2,\hbox{~} \theta^{B}_0\!=\!\epsilon,\hbox{~} \theta^{B}_1\!=\!-\pi/4,\hbox{~} 2\chi\!=\!\pi/4,
\end{eqnarray*}
where $\epsilon$ is a small angle ($0\!<\!\epsilon\!<\!\pi/40$) to ensure
that Eq.\ (\ref{eq: Condition for uniqueness}) only has a trivial solution.
As $\mathbf{P}$ and $\mathbf{Q}$ saturate
Eq.\ (\ref{eq: TLM inequality}) for scaled correlators,
they are the extremal points of ${\cal Q}_D$. 
Let us then consider $\mathbf{L}$ extrapolated from $\mathbf{P}$ and
$\mathbf{Q}$ as
\begin{equation}
\mathbf{P}=\lambda \mathbf{Q}+(1-\lambda)\mathbf{L},
\label{eq: Extrapolation}
\end{equation}
where $\lambda$ is chosen such that
$C_{00}\!+\!C_{01}\!+\!C_{10}\!-\!C_{11}\!=\!2$ at $\mathbf{L}$.
Suppose that $\{C^{A}_x,C^{B}_y,C_{xy}\}$
is extrapolated by Eq.\ (\ref{eq: Extrapolation}).
Because the behavior of $\mathbf{L}$ in the $C$-space satisfies the
positivity constraint
$p(ab|xy)\!\ge\!0$, $\mathbf{L}$ is a local correlation.
This implies that $\mathbf{P}$ can also be realized by a convex sum
of $\mathbf{Q}$ and deterministic correlations, despite that
$\mathbf{P}$ is an extremal point of ${\cal Q}_D$.
On the other hand, when $\{\delta^{B}_x,\delta^{A}_y,C_{xy}\}$
is extrapolated by Eq.\ (\ref{eq: Extrapolation}),
$\mathbf{L}$ is not allowed in quantum mechanics
as shown in Fig.\ \ref{fig: PQL}.
This implies that $(D_{x}^B)^2$ and $(D_{y}^A)^2$ must be strictly convex for
Eq.\ (\ref{eq: Extrapolation}).
Although it is unknown that this convex-sum realization certainly realizes
$\{\delta^{B}_x,\delta^{A}_y,C_{xy}\}$ of $\mathbf{P}$,
even an extremal point of ${\cal Q}_D$ may be realized as
a convex sum due to the convexity of $D_{x}^B$ and $D_{y}^A$.
Interestingly, as $\{C^{A}_x,C^{B}_y,C_{xy}\}$ of $\mathbf{P}$
in the $C$-space is realized by Eq.\ (\ref{eq: Extrapolation}),
$\mathbf{P}$ is not an extremal point of ${\cal Q}_C$, despite being an
extremal point of ${\cal Q}_D$.

%
\section{Uniqueness of geometry II}
\label{sec: Uniqueness of geometry II}

As a nonlocal behavior is considered, the measurement operators
in the realization satisfy $A_0\!\ne\!\pm A_1$ and $B_0\!\ne\!\pm B_1$
\cite{Fine82a}.
In the case of $\sin^22\chi\!=\!S^{p_{xy}}_{xy}\!=\!1$, the geometry of
real vectors is uniquely determined by the TLM criterion as shown
in \cite{Wang16a}. In the other cases, $\sin^22\chi\!=\!S^{p_{xy}}_{xy}$
is a solution of
\begin{equation*}
(\langle A_xB_y \rangle-\frac{\langle A_x\rangle\langle B_y\rangle}{\cos^22\chi})^2
=\sin^22\chi(1-\frac{\langle A_x\rangle^2}{\cos^22\chi})(1-\frac{\langle B_y\rangle^2}{\cos^22\chi}),
\end{equation*}
and $\langle A_x B_y\rangle$ is equal to either one of
\begin{equation}
\frac{\langle A_x\rangle\langle B_y\rangle}{\cos^22\chi}
\pm \sin2\chi\sqrt{1-\frac{\langle A_x\rangle^2}{\cos^22\chi}}\sqrt{1-\frac{\langle B_y\rangle^2}{\cos^22\chi}}.
\label{eq: C_xy}
\end{equation}
Let us introduce $\theta^{A}_x$ and $\theta^{B}_y$ by
\begin{equation}
\langle A_x\rangle=\cos2\chi \cos\theta^{A}_x,\hbox{~~}
\langle B_y\rangle=\cos2\chi \cos\theta^{B}_y.
\label{eq: A_x}
\end{equation}
Under this parameterization, 
\begin{equation}
d^{B}_x=\cos^22\chi\cos^2\theta^{A}_x+\sin^22\chi.
\label{eq: dbx}
\end{equation}
As $H\!\ge\!0$, the double sign of the second term in
Eq.\ (\ref{eq: C_xy}) can be negative for even pairs among the four possible
$(x,y)$, and hence,
by adjusting the sign of $\sin\theta^{A}_x$ and $\sin\theta^{B}_y$,
$\langle A_xB_y\rangle$ is always written as
\begin{equation}
\langle A_xB_y\rangle=\cos\theta^{A}_x\cos\theta^{B}_y
+\sin\theta^{A}_x\sin\theta^{B}_y\sin2\chi.
\label{eq: C_xy2}
\end{equation}
Let us then consider the real-vector representation.
Because the scaled correlators saturate Eq.\ (\ref{eq: TLM inequality}),
there exists real unit vectors $\vec F_x$ and $\vec E_y$ such that
\begin{equation}
\vec A_x\cdot\vec B_y=\sqrt{d^{B}_x}\vec F_x\cdot\vec B_y,
\hbox{~~}
\vec A_x\cdot\vec B_y=\sqrt{d^{A}_y}\vec E_x\cdot\vec A_x,
\end{equation}
and $\vec F_x$ and $\vec B_y$ ($\vec E_y$ and $\vec A_x$) are ensured to lie
in the same $B$-plane ($A$-plane) \cite{Wang16a}.
However, the relationship between the two planes
has not been determined yet.

Clearly, $\sqrt{d^{B}_x}$ is the length of the projection of $\vec A_x$ to
the $B$-plane, and from the laws of sines and cosines,
\begin{equation}
d^{B}_x=\frac{(\vec A_x\cdot\vec B_0)^2+(\vec A_x\cdot\vec B_1)^2
-2(\vec A_x\cdot\vec B_0)(\vec A_x\cdot\vec B_1)\cos\Delta}
{\sin^2\Delta},
\label{eq: Angle}
\end{equation}
must hold, where $\Delta$ is the angle between $\vec B_0$ and $\vec B_1$
(not yet determined).
From Eqs.\ (\ref{eq: dbx}) and (\ref{eq: C_xy2}), we can introduce
$\phi^{B}_x$ to express
$\vec A_x \cdot \vec B_y$ as $\sqrt{d^{B}_x}\cos(\phi^{B}_x\!-\!\theta^{B}_y)$,
and we have from Eq.\ (\ref{eq: Angle})
\begin{equation}
\big[\cos\Delta-\cos\Delta\theta^{B}\big]
\big[\cos\Delta-\cos(2\phi^{B}_x-\theta^{B}_0-\theta^{B}_1)\big]=0.
\end{equation}
As this must hold for both $x\!=\!0,1$, the solution of
$\cos\Delta\!=\!\cos(2\phi^{B}_x\!-\!\theta^{B}_0\!-\!\theta^{B}_1)$
is inappropriate unless the two-planes are perpendicular
(and the correlation is local).
We then have $\vec B_0\cdot\vec B_1\!=\!\cos\Delta\!=\!\cos\Delta\theta^{B}$.
Let the projector of $\vec \psi$ to the $B$-plane be $\vec\psi_{B}$.
As $\vec\psi\cdot\vec B_y\!=\!\langle B_y\rangle$,
\begin{equation}
|\vec\psi_B|^2=\frac{\langle B_0 \rangle^2+\langle B_1\rangle^2
-2\langle B_0\rangle\langle B_1\rangle
\cos\Delta\theta^{B}}
{\sin^2\Delta\theta^{B}}=\cos^22\chi,
\end{equation}
and hence we know from Eq.\ (\ref{eq: A_x}) that the angle between
$\vec \psi_B$ and $\vec B_y$ is $\theta^{B}_y$.
As $\vec \psi_B$ lies in the $B$-plane,
\begin{equation}
\vec\psi_B=\cos2\chi\frac{
\sin\theta^{B}_0 \vec B_1-\sin\theta^{B}_1\vec B_0}
{\sin\Delta \theta^{B}},
\end{equation}
and from Eq.\ (\ref{eq: C_xy2}) we have
$\vec A_x\cdot\vec \psi_B\!=\!\cos2\chi\cos\theta^{A}_x$,
which implies that the angle
between $\vec A_x$ and $\vec \psi_B$ is $\theta^{A}_x$.
From the same argument as above, we have
$\vec A_0\cdot\vec A_1\!=\!\cos\Delta\theta^{A}$,
which implies that $\vec A_0$, $\vec A_1$, and $\vec \psi_B$ lie in the
same plane.
Similarly, we know that $\vec B_0$, $\vec B_1$, and $\vec \psi_A$ lie in the
same plane.
After all, the geometry of real vectors is determined as
Fig.\ \ref{fig: Geometry} with $|\psi'|\!=\!\cos2\chi$.
The obvious symmetry is
$\{\theta^{A}_x,\theta^{B}_y,\chi\}$ and
$\{-\theta^{A}_x,-\theta^{B}_y,\chi\}$, which arises from the ambiguity
in adjusting the sign of $\sin\theta^{A}_x$ and $\sin\theta^{B}_y$.

In this way, without any assumption, the geometry is determined;
hence it is unique.
In the special case where
$S^{+}_{00}\!=\!S^{+}_{01}\!=\!S^{+}_{10}\!=\!S^{+}_{11}$ and
$S^{-}_{00}\!=\!S^{-}_{01}\!=\!S^{-}_{10}\!=\!S^{-}_{11}$,
there seem to exist two possible choices for $\sin^22\chi$.
However, as this contradicts the uniqueness of the geometry,
some condition is not satisfied for either choice.
For example, the correlation of the Tsirelson bound, where
$C^{A}_x\!=\!C^{B}_y\!=\!0$ and $C_{xy}\!=\!(-1)^{xy}/\sqrt{2}$,
we also have $S^{-}_{00}\!=\!S^{-}_{01}\!=\!S^{-}_{10}\!=\!S^{-}_{11}\!=\!1/2$,
but $H\!<\!0$ for this choice.


%

\end{document}